\documentclass{iau} 

\usepackage{graphicx}
\usepackage{xcolor}
\usepackage[hyphens]{url}
\usepackage{hyperref,comment}
\hypersetup{
     colorlinks  = true,
     linkcolor   = blue, 
     anchorcolor = BrickRed,
     citecolor   = blue,
     filecolor   = blue,
     menucolor   = blue,
     runcolor    = cyan,
     urlcolor    = blue
}
\usepackage[authoryear]{natbib}

\usepackage{aas_macros}
\usepackage{float} 
\usepackage{placeins} 

\pubyear{2019}
\volume{355} 
\jname{The Realm of the Low-Surface-Brightness Universe}
\editors{D. Valls-Gabaud, I. Trujillo \& S. Okamoto, eds.}

\title[The Huntsman Telescope] 
{The Huntsman Telescope}

\author[Spitler, Longbottom \& Alvarado-Montes]   
{Lee R. Spitler$^{1,2}$
 , Fergus D. Longbottom$^{1,2}$
 \mbox{Jaime A. Alvarado-Montes}$^{1,2}$,
 Amir E. Bazkiaei$^{1,2,3}$,
 Sarah E. Caddy$^{1,2}$,
 Wilfred T. Gee$^{1,2}$,
 Anthony Horton$^{4}$,
 Steven Lee$^{5}$
  \and 
 Daniel J. Prole$^{1,2}$
 }

\affiliation{$^1$Research Centre for Astronomy, Astrophysics \& Astrophotonics, Macquarie University, Sydney, NSW 2109, Australia\\
$^2$Department of Physics \& Astronomy, Macquarie University, Sydney, NSW 2109, Australia
$^3$Australian Research Council Centre of Excellence for All Sky Astrophysics in 3 Dimensions (ASTRO 3D)
$^4$Australian Astronomical Optics, Macquarie University, Sydney, NSW 2109, Australia\\
$^5$Research School of Astronomy and Astrophysics
Australian National University, Canberra ACT 2611, Australia}

\begin{document}

\maketitle

\begin{abstract}

The Huntsman Telescope, located at Siding Spring Observatory in Australia, is a system of ten telephoto Canon lenses designed for low surface brightness imaging in the Southern sky. Based upon the Dragonfly Telephoto Array, the refractive lens-based system provides an obstruction free optical path, which reduces the number of scattering surfaces and allows easier access to lower surface brightness levels.

In this proceeding, we present an analysis of the impact of flat fielding uncertainty on the limiting low surface brightness levels. We show that a fairly standard set of flat-field data can be well-characterised to a $\sim0.1\%$ level. This corresponds to a 5-$\sigma$ lower limit of $\sim33$ magnitude per arcsecond$^2$, which means that flat fielding is not likely going to set Huntsman's low surface brightness limit.

We also present early results of an exoplanet transient mode for Huntsman where all lenses work together to detect subtle variations in the luminosity of relatively bright $V=8-12$ magnitude stars. High-precision exoplanet imaging is ultimately limited by systematic uncertainties, so we anticipate multiple lenses will help to mitigate issues related to pixel-to-pixel and intra-pixel sensitivity variations. Our initial results show we can easily get $\sim0.4\%$ photometric precision with a single, defocused lens.

\end{abstract}

\firstsection 

\section{Introduction}

The motivation for low surface brightness imaging at optical wavelengths is covered by many of the other contributions in this Proceedings (e.g. Valls-Gabaurd 2019; Grebel 2019; Malin 2019; Mihos 2019). The limit of low surface brightness is normally set by a systematic uncertainty due to imperfections in flat-fielding, sky subtraction and removal of light from bright astronomical sources. While some of these can be mitigated with clever observation scheduling \citep{Trujillo_2016} one can also use a dedicated imaging system that is optimised for low surface brightness imaging \citep[e.g. ][]{Mihos2005,Abraham2014}.

The Huntsman Telescope\footnote{\url{http:\\\\huntsman.space}} is a Southern hemisphere version of the successful Dragonfly Telephoto Array \citep{Abraham2014}, a telescope system that consists of 48 off-the-shelf Canon lenses. Dragonfly has made a number of contributions to help understand galaxy evolution and formation \citep[Merritt this proceeding; Lokhorst this proceeding]{merritt_discovery_udg_2014,van_dokkum_forty-seven_2015}.

At the final stages of commissioning, Huntsman will operate in a fully automated mode with 10 Canon lenses. Each lens points at the same patch of sky and covers a field of view of $\sim3\times2$ degrees$^2$. The ultimate photometric limit for Huntsman will likely set by systematic uncertainties, thus work to date has focused on understanding those limits. Below are results of an investigation into systematic uncertainties due to flat fielding calibration and the feasibility of high-precision exoplanet photometry as a complementary science objective.

\section{Flatfielding}

\begin{figure}
 \centering
   \includegraphics[width=0.6\textwidth]{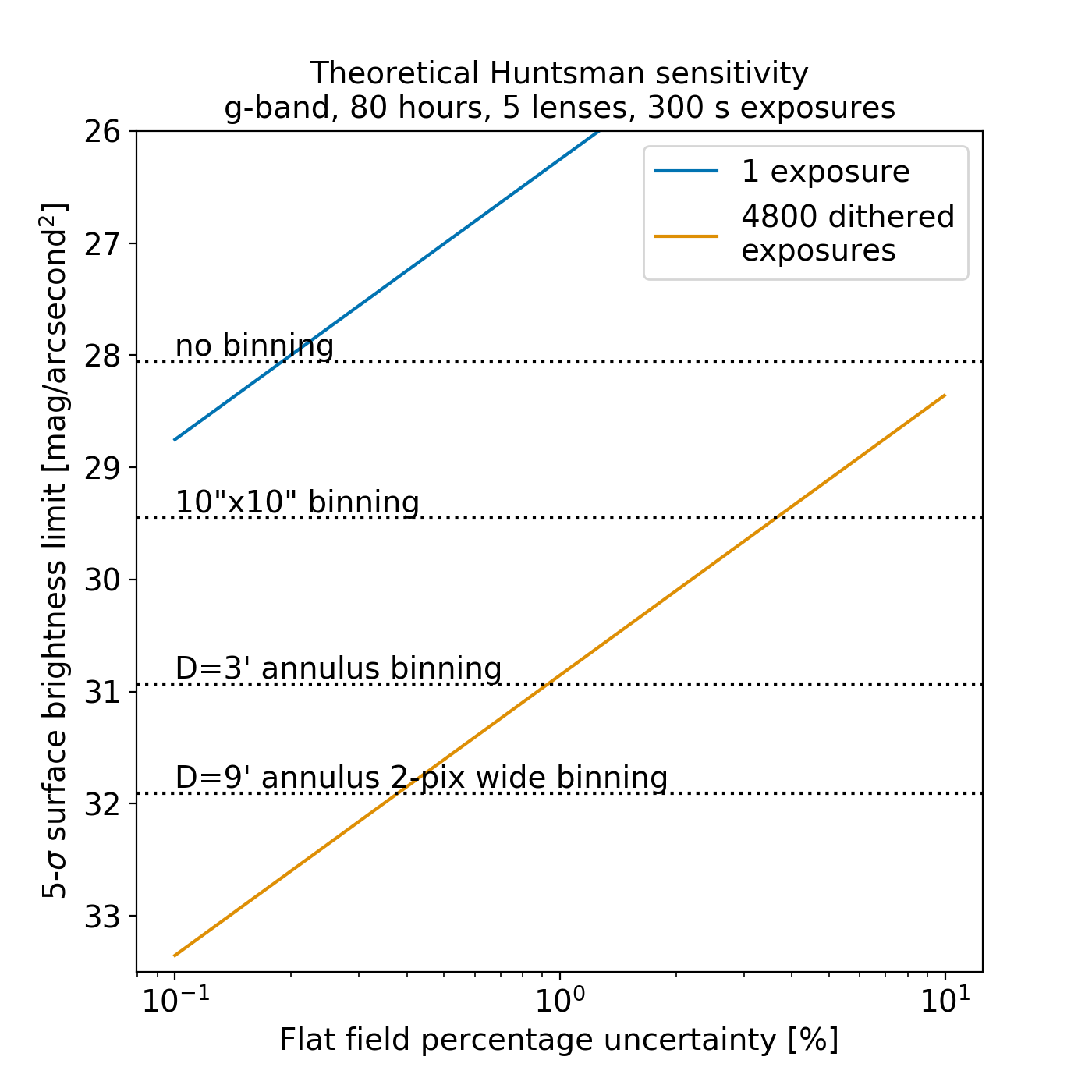}
   \caption[The impact of flat-field uncertainty on surface brightness depth limit.]{The theoretical achievable surface brightness depth limit when observations are limited by flat-fielding uncertainty given as percents on the horizontal axis. Various binning options are presented as dotted lines. Orange line indicates a full dataset for a single target: 80 hours with 5 lenses and 5-minute subexposures.}
   \label{fig:sbdepth} 
\end{figure}

A key innovation for Huntsman's potential as an instrument optimised for low surface brightness imaging are the unobstructed optical paths of its refracting lenses. The relatively fewer number of strong scattering surfaces between the incoming light and the CCD allows Huntsman to reduce systematic uncertainties as a result of scattered light and more easily reach lower surface brightness limits. This does not mean Huntsman is free of such issues; as the surface brightness floor is pushed lower, new sources of error become significant. One source of systematic error is flat fielding, as illustrated in \autoref{fig:sbdepth}. Simple propagation of a range of assumed flat fielding uncertainties through realistic observing campaigns and binning combinations, indicates they need to be smaller than 0.1\% to reach interesting surface brightness levels of $\sim\ 30\ mag\ arcsec^{-2}$ \citep[e.g. ][]{cooper2010}. These predictions were made with \texttt{Gunagala}\footnote{\url{https://github.com/AstroHuntsman/gunagala}}, a flexible python exposure time calculator.

Typically flat field data is collected using observations of the twilight sky or the uniformly illuminated inside wall of the telescope's dome. Within the literature, several authors have noted that dome flats were not sufficient for low surface brightness work, due to the difficulty in managing scattered light within the dome \citep{Feldmeier2002,Feldmeier2004,Trujillo2016}. Concerns have also been raised about differences in the colour of the twilight sky and the night sky \citep{Feldmeier2002,Feldmeier2004}, ruling out the use of twilight flats for some. As a result a majority of LSB flat-fielding seems to have been carried out using night sky flats, constructed from individual dark sky observations or from the actual science frames.

For the Huntsman Telescope, the combination of faint night sky levels in typical science data plus need for aggressive masking of the relatively large pixels means the commissioning datasets are not sufficiently large to construct high-quality flat field calibrations. As a result, this work focuses on using morning and evening twilight flats to assess the surface brightness limit set by flat fielding uncertainties.

Twilight flats present their own set of issues, as the twilight sky is not uniform over the $2\times3$ degree field of view for Huntsman. As reported by \citet{gradients}, these twilight gradients are relatively stable until late twilight and can be minimised by pointing towards a null point close to zenith, offset towards the anti-solar horizon. As these gradients will be aligned with the position of the rising or setting sun, it is possible to combine a set of morning and evening flats in order to remove the twilight gradient. This solution only works if both evening or morning flats are acquired in a given session. Without a set of both, it is much harder to remove the gradient due to the asymmetric vignetting pattern produced by the lenses.

\begin{figure}
 \centering
   \includegraphics[trim={0 0 0 9.24cm},clip,width=1.0\textwidth]{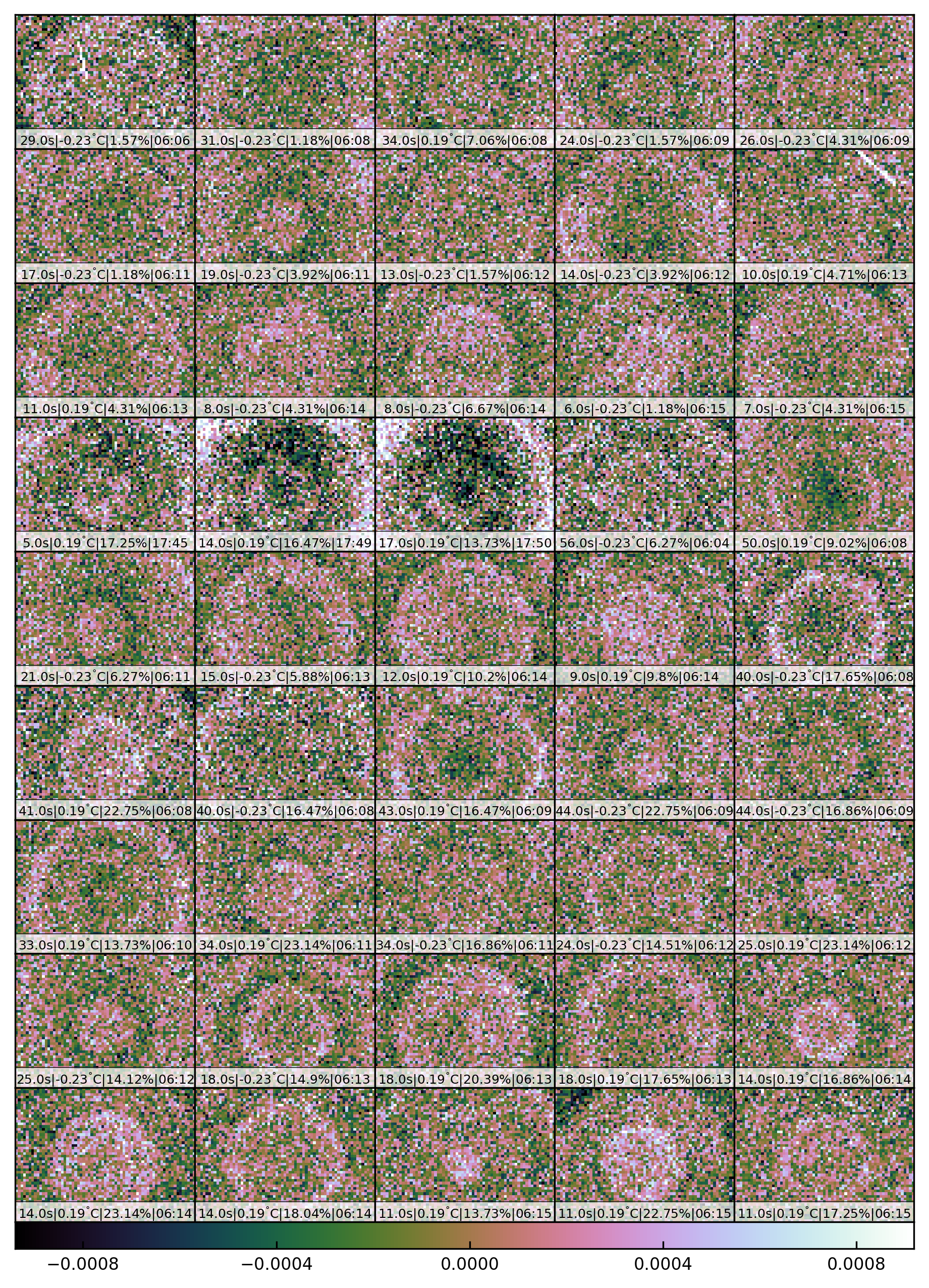}
   \caption[Residual circular features found in g-band twilight flats, sorted by local observing time.]{From \cite{Longbottom2019}, a selection of g-band individual flats after all predictable patterns (i.e. night sky gradients and the strong lens optical vignetting) sorted by local observing time. Noticeable in all panels is the presence of a centred circular or ring like structure(s) that varies in size and number. At the bottom of each panel the exposure time, CCD temperature and cooling percentage is given. Colorbar indicates fractional variations.}
   \label{fig:circles} 
\end{figure}

To quantify the intrinsic uncertainty of a single flat field frame, individual morning and evening twilight flats were divided by the average of median stacked morning and evening exposures. The flat field pattern is dominated by a circular optical vignetting pattern from the Canon lenses, where throughput falls by $\sim30\%$ from the centre to the edges of the field of view. After the strong vignetting is removed, a subtly-changing 0.8\% gradient across the CCD for individual flatfield exposures becomes apparent. The gradient is similar to that reported by \citet{gradients} and \citet{antarticflats}. After removing the 0.8\% gradients with a fitted plane, peculiar circular features at the $0.1\%$ level were found as shown in \autoref{fig:circles}.

The cause of the circular- or ring-like features is currently unknown. No apparent trend with wavelength, temperature, exposure time or time of night was found. If this feature cannot be mitigated, it would indicate the flat field calibration is good to a $0.1\%$ level. This means flat fielding is unlikely to set the low surface limit for Huntsman imaging (see \autoref{fig:sbdepth}). This is because $4800$ individual, dithered exposures will be combined from multiple lenses to make the final stacked science images.

\section{Exoplanets}

\begin{figure}
 \centering
   \includegraphics[width=0.95\textwidth]{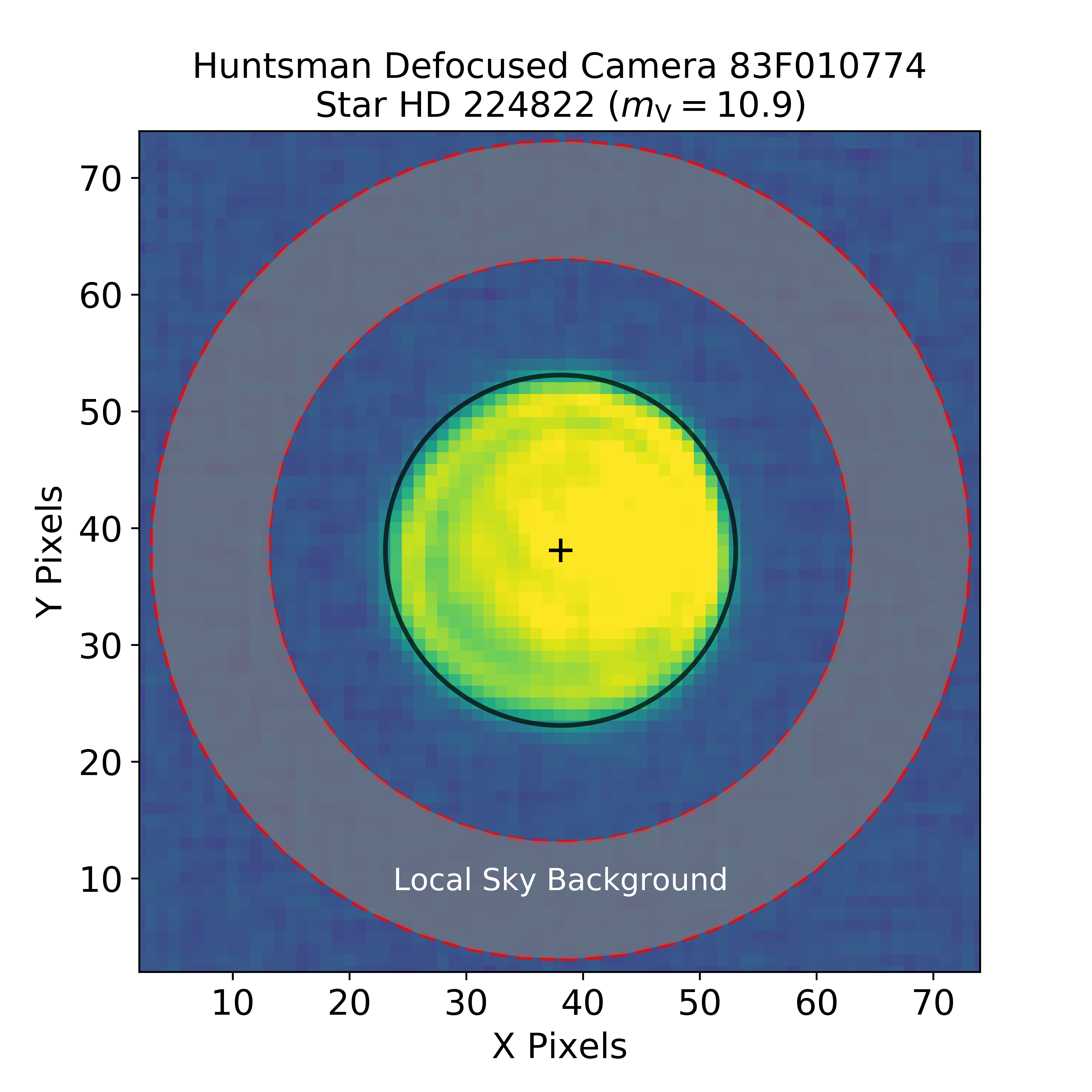}
   \caption[Index caption description.]{Circular aperture for the target star HD 224822 (black inner circle) and annular aperture for the local background approximation (red annulus and grey shadow).}
   \label{fig:figure4} 
\end{figure}

The Transiting Exoplanet Survey Satellite, TESS \citep{Ricker2015}, which started observing in August 2018, is scanning nearby stars and finding dozens of Earth- and Neptune-sized planet candidates \citep{Sullivan2015, Sullivan2017}. The number of unconfirmed exoplanet candidates from space-based missions will vastly surpass the follow-up capabilities on Earth so efficient and low-cost ground-based observatories are needed to identify the optimal targets for further exoplanet atmospheric characterisation using facilities like NASA's James Web Space Telescope, JWST \citep{Cowan2015,Batalha2017,Benneke2017}. Refractive lens systems consisting of a single lens paired with a high-end CCD have been shown to be very effective exoplanet transient detection systems \citep[e.g. ][]{Street2003, Christian2006,Guillon2008, Alonso2008,Zhao2014}. High-end CCDs have almost exclusively been used for this work to mitigate contributions from intra-pixel \citep{Mahato2018} and inter-pixel sensitivity variations \citep[e.g. ][]{Soutworth2009,Soutworth2009b,Mann2011,Croll2011a,Fukui2016} which impact photometric precision.

The purpose of this work is to explore the use of low-cost CCDs in combination with multiple lenses for exoplanet transient follow-up work. The low-cost CCDs are beneficial, if they can be shown to work, as they mean more telescope systems can be deployed. Another aspect of Huntsman that will be presented in a future work is showing how a multiple lens system can help reduce systematic uncertainties and achieve high-precision photometry.

Using data collected over a few nights with 4 lenses, we find that we can achieve high-precision photometry at a standard deviation of 0.4\% with minimal effort. A critical part of this is aggressive telescope defocusing (see \autoref{fig:figure4}) to spread light from a star over many pixels and avoid intra-pixel and inter-pixel sensitivity variations.

\begin{figure}
 \centering
   \includegraphics[width=0.95\textwidth]{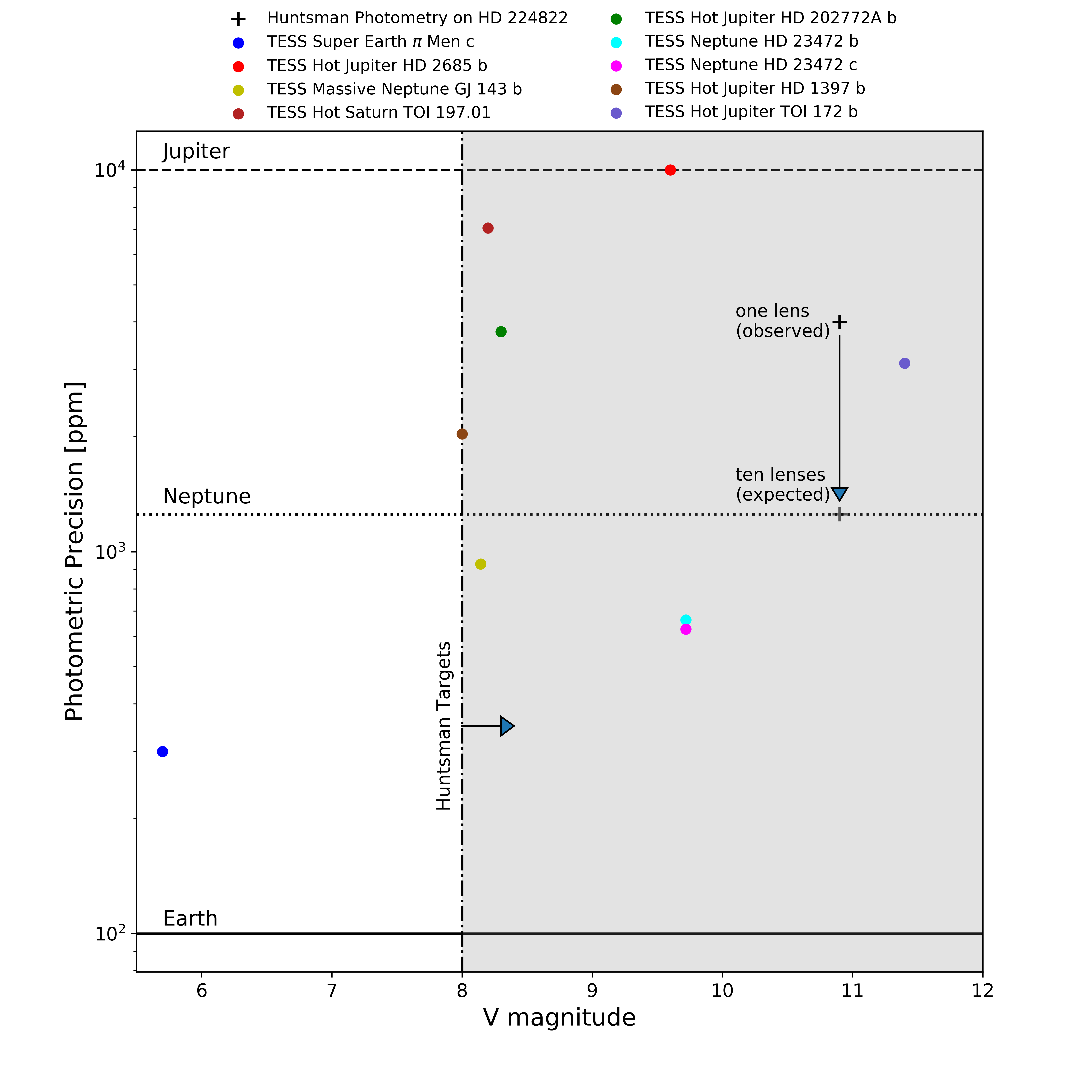}
   \caption[Photometric precision of Huntsman compared to known planets and exoplanets.]{This plot shows the photometric precision needed to detect any transit of confirmed TESS planets. Three different regimes are shown, namely, Jupiter-like, Neptune-like, and Earth-like planets (dashed, dotted, and solid horizontal lines respectively). The dashed-dotted vertical line stands for the target stars we will observe with Huntsman. The precision we found in \cite{Alvarado2019} when using one camera system for the target star HD 224822 is shown as a black cross. The expected precision for the future Huntsman array (10 lenses) is also shown. TESS-discovered planets include: $\pi$ Men c \citep{Huang2018}, HD 2685 b \citep{Jones2018}, HD 202772A b \citep{Wang2019}, HD 23472 b \citep{Trifonov2019}, HD 23472 c \citep{Trifonov2019}, GJ 143 b \citep{Trifonov2019}, TOI 197.01 \citep{Huber2019}, HD 1397 b \citep{Nielsen2019}, and TOI 172 b \citep{Rodriguez2019}. }
   \label{fig:TESShunt} 
\end{figure}

In \autoref{fig:TESShunt}, we present the precision found with Huntsman on the bright star HD 224822. For this we used only one camera and extrapolated the expected precision of ten lenses in the same figure. It is worth noting that there is still room for improvement -- the defocusing has not yet been optimised for Huntsman, so potentially a better precision can be achieved.

Planetary transits produce variations in the stellar flux of Sun-like stars from 0.01\% for Earth-like planets and 1\% for Jupiter-like planets. The photometric precision we get with Huntsman when using defocused photometry corresponds to Jupiter-like planets with a single lens. By binning over multiple exposures, and across the future Huntsman array of ten lenses, measuring the signal produced by Neptune-like planets will likely be possible. 

\FloatBarrier

\begin{acknowledgements}

LRS and DP acknowledge funding from a Australian Research Council Discovery Program grant DP190102448.

\end{acknowledgements}

\bibliographystyle{mnras}
\bibliography{refs_miscellaneous,refs_dragonfly,refs_big_picture,refs_data_reduction,refs_exoplanets}

\begin{discussion}
\discuss{J. Roman}{What is the precision of the flats made with the science images\,?}
\discuss{L. Spitler}{With limited science data we unfortunately cannot compute this yet.}

\discuss{I. Chilingarian}{How do you correct for the variable emission of the airglow, both in time and in space\,?}
\discuss{L. Spitler}{A combination of aggressive dithering and relatively long exposures will ensure the variations will cancel somewhat and can be modelled out.}
\end{discussion}

\end{document}